\documentclass[twocolumn, showpacs,aps,superscriptaddress,prl]{revtex4}

\usepackage{graphicx}
\usepackage{dcolumn}
\usepackage{bm}

\newcommand{\bB}{\mathbf{B}}
\newcommand{\eq}[1]{(\ref{#1})}

\begin{document}


\title{Scaling of spontaneous rotation with temperature and plasma current in tokamaks}

\author{F.~I.~Parra}
\email{fparra@mit.edu} \affiliation{Rudolf Peierls Centre for
Theoretical Physics, University of Oxford, Oxford OX1 3NP, UK}
\affiliation{Plasma Science and Fusion Center, Massachusetts
Institute of Technology, Cambridge, MA 02139, USA}
\author{M.~F.~F.~Nave}
\affiliation{Associa\c{c}\~{a}o EURATOM/IST, Instituto de Plasmas
e Fus\~{a}o Nuclear, Instituto Superior T\'ecnico, Technical University of Lisbon, Portugal}
\affiliation{Rudolf Peierls Centre for Theoretical Physics,
University of Oxford, Oxford OX1 3NP, UK}
\author{A.~A.~Schekochihin}
\affiliation{Rudolf Peierls Centre for Theoretical Physics,
University of Oxford, Oxford OX1 3NP, UK}
\author{C. Giroud}
\affiliation{Euratom/CCFE Fusion Association, Culham Science
Centre, Abingdon OX14 3DB, UK}
\author{J.~S.~de~Grassie}
\affiliation{General Atomics, P.O. Box 85608, San Diego,
California, USA}
\author{J.~H.~F.~Severo}
\affiliation{Institute of Physics, University of S\~{a}o Paulo,
SP, Brazil}
\author{P.~de~Vries}
\affiliation{FOM Institute for Plasma Physics, Rijnhuizen,
Association EURATOM-FOM, Nieuwegein, The Netherlands}
\author{K.-D.~Zastrow}
\affiliation{Euratom/CCFE Fusion Association, Culham Science
Centre, Abingdon OX14 3DB, UK}
\author{JET-EFDA Contributors \footnote{See the Appendix of F. Romanelli et al., Proceedings of the 23rd IAEA Fusion Energy Conference 2010, Daejeon, Korea.}}
\affiliation{JET-EFDA, Culham Science Centre, Abingdon, OX14 3DB, UK}

\date{\today}

\begin{abstract}
Using theoretical arguments, a simple scaling law for the size of the intrinsic rotation observed in tokamaks in the absence of momentum injection is found: the velocity generated in the core of a tokamak must be proportional to the ion temperature difference in the core divided by the plasma current, independent of the size of the device. The constant of proportionality is of the order of $10\,
\mathrm{km \cdot s^{-1} \cdot MA \cdot keV^{-1}}$. When the intrinsic rotation profile is hollow, i.e. it is counter-current in the core of the tokamak and co-current in the edge, the scaling law presented in this Letter fits the data remarkably well for several tokamaks of vastly different size and heated by different mechanisms.
\end{abstract}

\pacs{52.25.Fi, 52.30.-q, 52.55.Fa}
\maketitle

\emph{Introduction.} Due to their axisymmetry, tokamak plasmas can
be made to rotate at high speeds if momentum is injected into
them. If the rotation is sufficiently large, large scale magnetohydrodynamic (MHD) instabilities are stabilized \cite{deVries96} and  the turbulent
transport of energy can be much reduced \cite{connor04,
deVries09, mantica10}. Unfortunately, ITER \cite{ikeda07}, the largest
magnetic confinement experiment currently being built, is not
expected to have effective momentum deposition due to its size and
high density. As a result, there has been mounting interest in the
intrinsic, or spontaneous, rotation observed in tokamaks without
momentum injection \cite{rice07}. If this intrinsic rotation could be made
large, it could be used to prevent instabilities and reduce turbulence as is done with
momentum injection. Understanding the origin of this rotation is
also an interesting physics question. This has driven several experimental \cite{scarabosio06, duval07, rice07, degrassie07, eriksson09}, numerical \cite{wang10, camenen11, wang11} and theoretical \cite{diamond08, camenen09, parra10, parra11d} studies. So far, numerical results for intrinsic rotation have only been obtained using global gyrokinetic simulations that have been recently proven to be flawed for radial momentum transport in the core of tokamaks \cite{parra10b}.

In this Letter, we use very simple theoretical arguments to show that the velocity
difference within the core of a tokamak must scale proportionally
to the ion temperature divided by the plasma current. The constant of proportionality is independent of machine size and is of order $c^2/e = 10\, \mathrm{km \cdot s^{-1} \cdot MA \cdot
keV^{-1}}$, where $c$ is the speed of light and $e$ is the proton
charge. We show that pulses with hollow intrinsic rotation (counter-current rotation at the magnetic axis and co-current at the edge)  from machines whose sizes
range from tens of centimeters to several meters, that have very
different plasma currents (from 0.1 MA to 2.5 MA), and that are
heated by different mechanisms (JET \cite{eriksson09}, DIII-D
\cite{degrassie07}, TCABR \cite{severo03} and TCV
\cite{scarabosio06}) follow the theoretical scaling.

\emph{Theoretical arguments.} In a tokamak plasma, turbulence and
collisions transport momentum across
magnetic surfaces. Momentum can be injected with neutral beams and radio frequency waves (RF) \cite{incecushman09}, but in many occasions there is no external
source of momentum. When the latter is the case, the toroidal
angular momentum flux $\Pi$ through every flux surface must be
zero, even though significant rotation can often be observed experimentally. Only the angular momentum in the toroidal direction is
relevant. In the poloidal direction, the flow is strongly damped
by collisional processes, which pass the momentum through the
magnets to the structure of the tokamak. Thus, to calculate
intrinsic rotation profiles, it is necessary to calculate the
dependence of $\Pi$ on the toroidal rotation frequency
$\Omega_\phi$ and then solve the equation $\Pi (\Omega_\phi) = 0$
for $\Omega_\phi$.

Both turbulence and collisions occur on time scales that are
longer than the inverse of the gyrofrequency, which means that the
particle trajectories can be understood as a fast gyromotion
around guiding centers, which move fast along magnetic field lines
and drift slowly across them. This is the physical idea underlying
gyrokinetics, which is the most commonly used approximation in
transport simulations \cite{dorland00, candy03, dannert05,
peeters09, barnes10}.

Even in the absence of turbulence and collisions, particles move
out of the surface of constant magnetic flux where they started
due to the $\nabla B$ and curvature drifts, but they remain within
a given distance of it. This distance is of the order of
the poloidal gyroradius $\rho_\theta = m c v_{th}/e B_\theta$,
where $e$ and $m$ are the charge and mass of the particle,
$v_{th}$ is the thermal speed, and $B_\theta$ is the poloidal
component of the magnetic field. Note that $\rho_\theta =
(B/B_\theta) \rho$, where $\rho$ is the particle gyroradius and
$B$ is the total magnetic field. In most tokamaks, $B/B_\theta$ is
of order $10$. Tokamaks are constructed so that $\rho_\theta \ll
L_T$, where $L_T$ is the characteristic length of variation of the
temperature $T$, which we use as our length of reference.

Collisions cause transport, known as neoclassical transport \cite{hinton76}, because each collision makes the particle move from one drift orbit to another separated by $\rho_\theta$. Turbulent transport is caused by electromagnetic fluctuations, of which the most virulent are believed to be driven by the ion temperature gradient (ITG). For ITG turbulence well above marginality, the characteristic correlation length is $(B/B_\theta) (a/L_T) \rho \sim \rho_\theta$, where $a$ is the minor radius of the tokamak ($L_T \sim a$). This scaling is not based on the drift orbits as is in the case of collisional transport, but on critical balance between the parallel and the perpendicular dynamics \cite{barnes11b}. It is observed in experiments that the turbulent transport scales approximately linearly with $B/B_\theta$ \cite{petty04}, as predicted by critical balance \cite{barnes11b}

In general, tokamaks are geometrically up-down symmetric to a
great degree in the core. In such tokamaks, to lowest order in
$\rho_\theta/L_T$, the transport of momentum can only be different
from zero if a preferred direction is given by either rotation or
rotational shear. The lowest order cancellation of the radial momentum flux in the absence of rotation is due to a fundamental symmetry of the turbulence and the particle motion \cite{peeters05, sugama11, parra11}. Here we are assuming $\Omega_\phi \sim v_{th}/R$, this being the ordering for which the rotation and its shear enter in the lowest order gyrokinetic equation \cite{artun94, sugama98}. Thus,
schematically, to lowest order in $\rho_\theta/L_T$,
\begin{equation} \label{sketchmomentum}
\Pi \sim - \nu_\mathrm{t} R^2 \left ( \frac{\partial
\Omega_\phi}{\partial r} + \frac{\Omega_\phi}{\ell_\mathrm{pinch}}
\right ) - \nu_\mathrm{c} R^2 \frac{\partial \Omega_\phi}{\partial
r},
\end{equation}
where $r$ is the radial coordinate, $R$ is the major radius,
$\nu_\mathrm{t}$ is the turbulent viscosity, $-
\nu_\mathrm{t}/\ell_\mathrm{pinch}$ is the turbulent pinch of
momentum \cite{peeters07, tardini09}, and $\nu_\mathrm{c}$ is the
collisional viscosity. Equation \eq{sketchmomentum} has the main features of momentum transport in up-down symmetric tokamaks: momentum transport can only happen when $\Omega_\phi \neq 0$ or $\partial \Omega_\phi/\partial r \neq 0$, and it changes sign when $\Omega_\phi$ and $\partial \Omega_\phi/\partial r$ do \cite{parra11}. It can be thought of as a Taylor expansion of the complicated function $\Pi (\partial \Omega_\phi/\partial r, \Omega_\phi)$ around $\Omega_\phi = 0$ and $\partial \Omega_\phi/\partial r = 0$.

The equation for intrinsic rotation is $\Pi = 0$, and with the
lowest-order expression \eq{sketchmomentum} for $\Pi$, the
solution is $\Omega_\phi \propto \exp [ - \int
dr\,\ell_\mathrm{pinch}^{-1} (1 + \nu_\mathrm{c}/\nu_\mathrm{t})^{-1} ]$. It is then possible to obtain intrinsic rotation if rotation is generated in some region of the plasma
(for example, in the edge) and pinched to other regions. However,
this mechanism is not fully satisfactory because it cannot explain
the variety of observed profiles \cite{eriksson09}. In particular,
$\Omega_\phi$ cannot change sign, contradicting experimental
observations (as we will show in the next section). Unfortunately,
to lowest order, Eq.~\eq{sketchmomentum} is correct and no other
mechanism for intrinsic rotation can be obtained.

\begin{figure}
\includegraphics[width = 7.5 cm, height = 5.8cm]{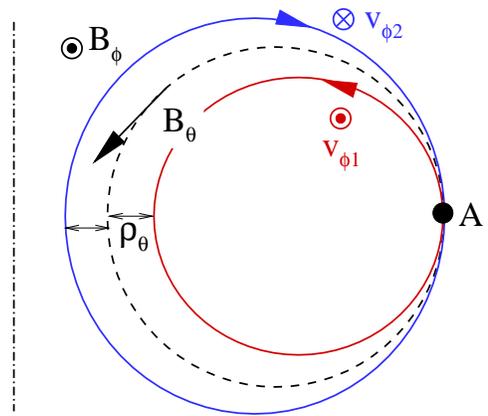}

\caption{\label{sketch_orbit} Sketch of drift orbits.}
\end{figure}

If the expansion in $\rho_\theta/L_T \ll 1$ is continued to next
order, the rotation and its shear are not the only physical
factors that provide a preferred direction and can give raise to
momentum transport: the pressure and temperature gradients also
break the up-down symmetry. Consider the guiding centers of two particles
(1 and 2) that at point A at the outboard midplane of a
tokamak have velocities in opposite directions, as sketched in
Fig.~\ref{sketch_orbit}. The dashed line represents the cut of a
surface of constant magnetic flux through a poloidal plane (the
axis of symmetry is the dash-dot line). The poloidal magnetic
field $B_\theta$ is parallel to the dashed line and points
counterclockwise, whereas the toroidal magnetic field $B_\phi$
points towards the reader. At point A, particle 1 (red orbit)
travels counterclockwise, and since to lowest order it follows the
magnetic field, its toroidal velocity $v_{\phi 1}$ is pointing
towards the reader. Particle 2 (blue orbit) travels in the
opposite direction both poloidally and toroidally. Orbits separate from the flux surface a small
distance of order $\rho_\theta$. Particle 1 moves towards the
center of the tokamak because its poloidal velocity is counterclockwise. Particle 2 drifts
outwards. Because of the temperature gradient, the center of the
tokamak is hotter, and particles like particle 1 will have more
energy, of the order of $(\rho_\theta/L_T) m v_{th}^2$, breaking
the symmetry and, in this simplified picture, making the plasma
rotate counterclockwise poloidally, and towards the reader
toroidally. Fig.~\ref{sketch_orbit} shows that whereas the
direction of the magnetic field is unimportant, the vector $\bB
\times \nabla T$ does give a preferred direction at higher order
in $\rho_\theta/L_T$ parallel to or against which the plasma will
tend to rotate. The mechanism described here does not determine
the sense of the toroidal rotation, but it does demonstrate that
background gradients break the up-down symmetry and that the
effects of this symmetry breaking are of order $\rho_\theta/L_T$.
Calculating all these effects is a rather sophisticated analytical
task, involving many factors subtler than the simple argument
given above \footnote{To give an idea of the complexity of the
theoretical problem, we point out that the poloidal velocity will
not be counterclockwise at low collisionality, as suggested by
Fig.~\ref{sketch_orbit}, but clockwise due to the collisional
friction between particles of the type sketched in
Fig.~\ref{sketch_orbit} and a population of trapped particles
\cite{hinton76}. Determining the toroidal rotation requires even
more work than the poloidal velocity, and the finite-orbit-width
effects come in through other mechanisms. For example, the
poloidal velocity breaks the up-down symmetry and gives a
preferred direction to the turbulent and collisional transport of
toroidal angular momentum.}.

The next-order contributions to momentum transport in
$\rho_\theta/L_T \ll 1$ were first calculated in neoclassical
theory \cite{catto05, wong09}, where they are proportional to
radial derivatives of the ion temperature. Models to calculate the
next-order contributions to turbulent transport have also been
proposed \cite{parra10, parra11d}. Near marginality, the turbulence amplitude is small, and the neoclassical corrections to the distribution function of order $\rho_\theta/L_T$ due to finite drift orbit size are the dominant mechanism that breaks the up-down symmetry of the turbulence. Well above marginality, the characteristic eddie size is $\rho_\theta$ \cite{barnes11b}, allowing the turbulence to sample regions in which the temperature gradient differs by $\rho_\theta/L_T$, and breaking the symmetry this way. In general, we expect the new
next-order terms to depend strongly on density and temperature
gradients because these drive the turbulence. Schematically, as shown in \cite{parra11d}, we
may write
\begin{eqnarray} \label{intrinsicmomentum}
\Pi \sim - \nu_\mathrm{t} R^2 \Bigg [ \frac{\partial
\Omega_\phi}{\partial r} + \frac{\Omega_\phi}{\ell_\mathrm{pinch}}
+ O \left ( \frac{\rho_\theta}{L_T} \frac{v_{th}}{R L_T} \right )
\Bigg ] \nonumber\\- \nu_\mathrm{c} R^2 \left [ \frac{\partial
\Omega_\phi}{\partial r} + O \left ( \frac{\rho_\theta}{L_T}
\frac{v_{th}}{R L_T} \right ) \right ].
\end{eqnarray}
From \eq{intrinsicmomentum}, setting $\Pi = 0$ and assuming that
the scale length of $\Omega_\phi$ is of order $L_T$, we obtain $R
\Omega_\phi \sim (\rho_\theta/L_T) v_{th} \sim (c/e B_\theta)
(T/L_T)$, where $T = m v_{th}^2/2$ is the temperature. The
poloidal magnetic field is given by the toroidal plasma current
$I_p$, $B_\theta \sim I_p/c L_B$, where $L_B$ is the
characteristic length of variation of $B_\theta$. Therefore, $R
\Omega_\phi \sim (L_B/L_T) (c^2/e) (T/I_p)$. In the core, $L_B$
and $L_T$ are both of the order of the minor radius $a$, so the
toroidal velocity is
\begin{equation} \label{scaling}
V_\phi = R \Omega_\phi \sim \frac{c^2}{e} \frac{T}{I_p}.
\end{equation}
This equation gives the scaling of intrinsic rotation in the core with temperature and plasma current.
It is \emph{independent of machine size.} The dimensional constant
of proportionality is $c^2/e = 10\, \mathrm{km \cdot s^{-1} \cdot
MA \cdot keV^{-1}}$.

\begin{figure}
\includegraphics[width = 8.5 cm, height = 5.3cm]{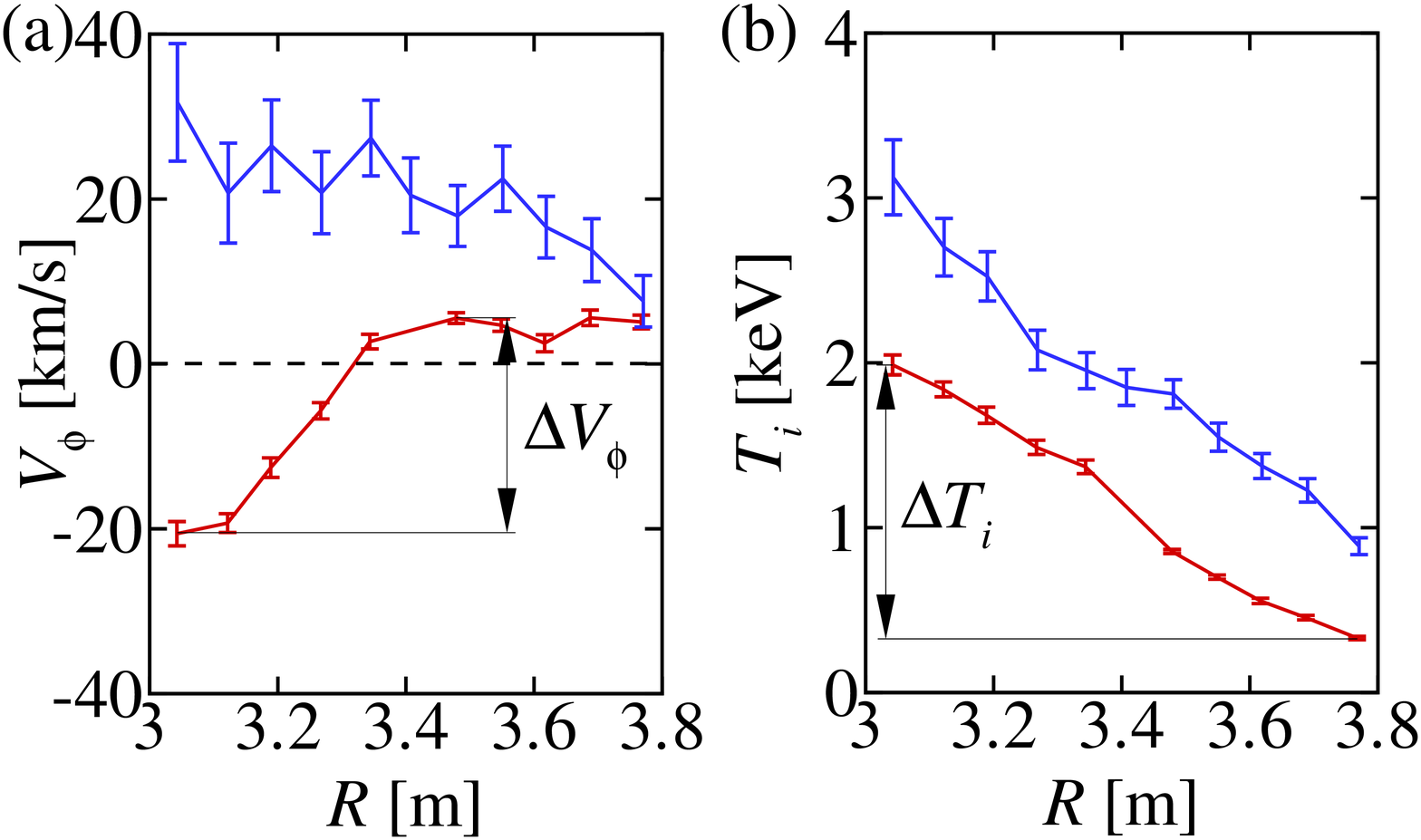}

\caption{\label{profiles} Intrinsic rotation profiles (a) and ion
temperature profiles (b) in JET plasmas with ICRH, pulses 66395
(red) and 74692 (blue). The rotation in the co-current direction
is positive rotation. The position of the magnetic axis is around
$R = 3\, \mathrm{m}$, the separatrix is around $R = 3.8\,
\mathrm{m}$.}
\end{figure}

\emph{Experimental measurements.} We now compare experimental data
from different machines that show similarities in their intrinsic
rotation profiles. In Fig.~\ref{profiles} two pulses from
JET represent two distinct types of intrinsic rotation
profiles: the ones in which the toroidal velocity increases from
the magnetic axis towards the edge of the tokamak (red profile),
which we call \emph{hollow profiles}, and the ones in which it
decreases (blue profile), which we call \emph{peaked profiles}
(the toroidal velocity is deemed positive if it is co-current).
The two pulses in Fig.~\ref{profiles} have very different input
power and plasma current, and they are only meant to be examples
of the two types of velocity profiles. The peaked profiles need
not have higher temperature gradients than the hollow profiles.
The velocity at the edge is mostly co-current, and this seems to
be common to all tokamaks with low magnetic ripple in the absence of momentum injection. In JET, the hollow profiles correspond
to Ohmic shots and some of the Ion Cyclotron Resonance Heating
(ICRH) pulses in both Low-Confinement Mode (L-mode)
\cite{eriksson09} and High-Confinement Mode (H-mode)
\cite{nave11}. The cases with peaked profiles are all ICRH L-mode
and H-mode shots.

\begin{figure}
\includegraphics[width = 8.5 cm, height = 6.8cm]{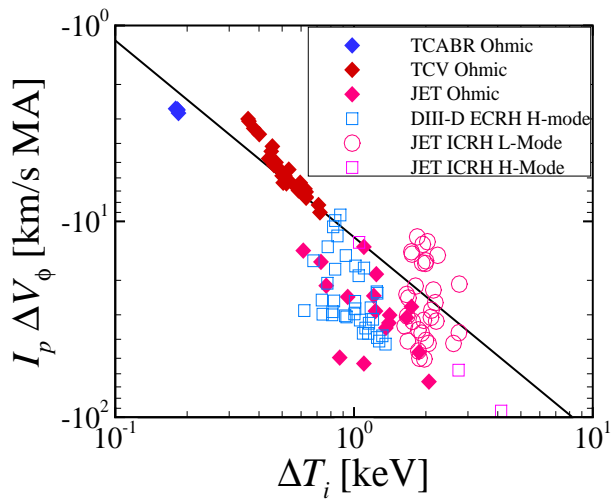}

\caption{\label{comparison} Toroidal velocity difference in the
core $\Delta V_\phi$ multiplied by plasma current $I_p$ against
the ion temperature difference $\Delta T_i$ in the plasma core.
The line is the least-square fit of the data to \eq{scalexp}. The
slope is $18\, \mathrm{km \cdot s^{-1} \cdot MA \cdot keV^{-1}}$.}
\end{figure}

To check \eq{scaling}, we compare the pulses with hollow core
velocity profile for four different tokamaks: JET
\cite{eriksson09}, DIII-D \cite{degrassie07}, TCABR
\cite{severo03} and TCV \cite{scarabosio06}. To characterize the
velocity generated intrinsically in the core, we use the
difference in toroidal velocity $\Delta V_\phi$ between the
minimum of toroidal velocity closest to the magnetic axis on the
outboard side and the first maximum encountered when moving from
the magnetic axis towards the edge on the outboard side. This
definition of $\Delta V_\phi$ is illustrated in
Fig.~\ref{profiles}(a). The parameter $\Delta V_\phi$ attempts to
exclude any intrinsic velocity generated at the edge -- most
likely by means not covered in our theoretical discussion above.
To give a measure of the sources generating intrinsic rotation in
\eq{intrinsicmomentum}, we use the difference in ion temperature
$\Delta T_i$ between the magnetic axis and the temperature at the
top of the pedestal in H-modes, or the temperature measurement that is the closest to the
separatrix in L-modes. The difference $\Delta T_i$, illustrated in
Fig.~\ref{profiles}(b), excludes the ion temperature jump in the
pedestal in the case of H-modes. Fig.~\ref{comparison} shows
$I_p \Delta V_\phi$ vs. $\Delta T_i$ for various tokamaks
\footnote{The data for TCV was obtained from \cite{scarabosio06}.
In that article, the authors plotted the velocity at the magnetic
axis multiplied by $I_p$ against the temperature at the magnetic
axis. This is approximately the same as $I_p \Delta V_\phi$ vs.
$\Delta T_i$ because both the rotation and the temperature at the
edge are small compared to their values at the magnetic axis.}.
According to \eq{scaling}, we expect
\begin{equation} \label{scalexp}
I_p \Delta V_\phi = \alpha \frac{c^2}{e} \Delta T_i.
\end{equation}
The dimensionless prefactor $\alpha$ could not be determined in
our qualitative theoretical discussion, but we can find its
value from the present experimental analysis. The data is
consistent with an approximate linear dependence with a slope of
$(18 \pm 4)\, \mathrm{km \cdot s^{-1} \cdot MA \cdot keV^{-1}}$
for all machines, giving $\alpha \simeq 1.8 \pm 0.4$. The slope
was determined by least-square fitting and the error is the $99\%$
confidence interval. In Fig.~\ref{comparison} there are
Ohmic, ICRH and Electron Cyclotron Resonance Heating (ECRH) shots,
L- and H-modes, plasma currents spanning from 0.1 MA to 2.5 MA,
and machines of sizes ranging from tens of centimeters (TCABR) to
meters (JET). The fact that both the scaling and the prefactor
seem to be valid for this variety of situations suggests that the
theoretical ideas proposed above are robust.

When the same analysis was attempted for the peaked profiles in JET, the trend was not so clear, but we cannot conclude that the scaling is absent either. The study of peaked profiles will be the object of a future publication. Here we review several possible explanations for the lack of a clear scaling. In \cite{duval07, camenen11} a change from Trapped Electron Mode (TEM) driven turbulence to ITG turbulence was proposed as the cause for the transition between peaked and hollow profiles. If this is the case, $\Delta T_i$ is not a good parameter to work with because TEM turbulence depends strongly on the electron density gradient, for which $\Delta T_i$ is not a good proxy, and on the electron temperature profile that for low collisionality may differ from $T_i$. It is also possible that the peaked-profile cases are dominated by the inward pinch of momentum generated at the edge \cite{eriksson09}, making the rotation in the core correlated to the parameters at the edge and not to the parameters of the core. With our preliminary analysis of peaked profiles in JET, we cannot decide if the transition from hollow to peaked profiles is due to the reasons given above, or other reasons not considered here. Experiments in DIII-D show that shaping affects intrinsic rotation, with high triangularity shots tending to have peaked profiles. Shaping may affect ITG and TEM turbulence differently, making one or the other type of turbulence dominant for high triangularity and hence deciding the direction of rotation in this way, or it may have an effect on the direction in which ITG or TEM turbulence pushes the plasma. Further study is needed. Even though the trend with $\Delta T_i$ and $I_p$ was not so clear, the velocity difference $\Delta V_\phi$ was still of the same order as \eq{scaling}.

\emph{Discussion.} Using simple theoretical arguments, we have shown that the intrinsic rotation
generated in the core must scale according to \eq{scalexp}.  Hollow intrinsic rotation profiles from very different tokamaks follow this scaling. The scatter in Fig.~\ref{comparison} is to be expected since \eq{scalexp} is derived from an order of magnitude estimate and prefactors of order unity may vary from shot to shot.

There are ways of generating intrinsic rotation that have not been
considered in this Letter. For example, in the core, RF heating
can transport momentum \cite{perkins01,
eriksson02} due to the large orbits of energetic ions. In the edge,
direct particle losses can generate rotation \cite{degrassie09a}. It seems that these
effects are not important in the cases presented in
Fig.~\ref{comparison} because these include shots with and without
energetic ions, and with and without a pedestal. We do not know
how generic this is. We have introduced a dimensionless parameter
$\alpha = e I_p \Delta V_\phi/c^2 \Delta T_i$, which was of order
unity for a variety of regimes and machines considered here. It
would be very instructive to quantify experimentally measured
rotation in other cases in terms of this parameter. In cases that
$\alpha$ is significantly larger than unity, the rotation must
have external origin, such as energetic ions, edge effects or
momentum injection.

The experimental results presented above cannot determine if the
transport of momentum is dominated by collisions or turbulence
because both have the same scaling \eq{scaling}. Since turbulent
viscosity is of the same order as the thermal diffusivity
\cite{tardini09, casson09, barnes11, highcock10}, and turbulent
transport usually dominates, we expect the $\rho_\theta/L_T$
corrections to the turbulent momentum transport to play the
dominant role in driving intrinsic rotation.

The authors acknowledge many helpful discussions with M. Barnes
and T. Johnson. This work was supported in part by EPSRC, STFC, Funda\c{c}\~{a}o para a Ciencia e
Tecnologia (FCT) of Portugal,
the Leverhulme Trust Network for Magnetized Plasma Turbulence, the
US DOE under DE-FC02-04ER54698, the European Communities through
Association Euratom/IST, and carried out within the frameworks of
the Instituto de Plasmas e Fus\~{a}o Nuclear-Laboratorio Associado
and the European Fusion Development Agreement. The views and opinions expressed
herein do not necessarily reflect those of the European
Commission.

\end{document}